\documentclass[baaa]{baaa}

\usepackage[pdftex]{hyperref}
\usepackage{subfigure}
\usepackage{natbib}
\usepackage{helvet,soul}
\usepackage[font=small]{caption}

\contriblanguage{0}

\contribtype{1}

\thematicarea{7}

\title{Estrellas híbridas con núcleos superconductores de color}

\titlerunning{Estrellas híbridas}

\author{D. Curin\inst{1}, {I.F.} Ranea-Sandoval\inst{1,2} , M. Orsaria\inst{1,2} {\&} {G.A.} Contrera\inst{1,2,3}}
\authorrunning{Curin et al.}

\contact{danielacurin@fcaglp.unlp.edu.ar}

\institute{
Grupo de Gravitación, Astrofísica y Cosmología, Facultad de Ciencias Astronómicas y Geofísicas, UNLP, Argentina 
\and
Consejo Nacional de Investigaciones Científicas y Técnicas, Argentina 
\and
Instituto de Física de La Plata, CONICET{--}UNLP, Argentina
}

\resumen{
El objetivo de este trabajo es el estudio del interior de estrellas híbridas a partir ecuaciones de estado que describen la materia que las compone. Usamos ecuaciones de estado hadrónicas modernas para describir la materia del núcleo externo de estas estrellas. La fase de quarks es modelada por una versión extendida del modelo {\it Field Correlator Method}, en la cual se incluyen interacciones vectoriales entre los quarks y superconductividad de color. Analizamos el efecto de estas dos contribuciones en la relación masa-radio y en la deformabilidad por {mareas. Esta es} una cantidad que se volvió relevante para las estrellas de neutrones luego del evento GW170817, que impone restricciones adicionales a la ecuación de estado de estos objetos compactos.}

\abstract{
The aim of this work is the study of hybrid stars interiors from the equations of state describing the matter composing them. We used modern hadronic equations of state to describe the matter in the outer core of these stars. The quark phase is modeled by an extended version of the Field Correlator Method, in which vector interactions among quarks and color superconductivity are included. We analyze the effect of these two contributions on the mass-radius relationship and tidal {deformability. This is a} relevant quantity for neutron stars after {the} GW170817 event, {allowing} to impose additional constraints to the equation of state of these compact objetcs.}


\begin{document}

\maketitle

\section{Introducción}
\label{Intro}
Las estrellas de neutrones (ENs) son el remanente compacto de una estrella masiva ($>$~10~$\textrm{M}_{\odot}$) al final de su vida, luego de una explosión de Supernova tipo II \cite{Cerda_Duran_2018}.

La observación de los púlsares PSR J1614-2230 (\citep{2010Natur.467.1081D}, luego corregido por \citep{2018ApJS..235...37A}), y PSR J0348+0432  \cite{2013Sci...340..448A}, imponen un límite inferior de 2~$\textrm{M}_{\odot}$ para las masas máximas de las familias de ENs construidas con una ecuación de estado (EdE) determinada. Por otro lado, el evento conocido como GW170817, en el que se detectó una fusión de dos ENs, impone una restricción adicional para los radios de estrellas de 1.4~$\textrm{M}_{\odot}$ entre 9.6~km y 13.76~km \citep{2018PhRvC..98d5804T}.

Para describir la composición interna de estos objetos, en los que la materia está sometida a condiciones extremas de presión y densidad, se usan modelos efectivos que reproducen las características básicas de la QCD.
En este trabajo se proponen posibles EdE para la descripción de la materia dentro de una estrella híbrida (EH): estrella de neutrones con núcleo interno de materia de quarks y núcleo externo de materia hadrónica.
Luego, se analiza la estructura interna de las EHs resolviendo las ecuaciones de equilibrio hidrostático, conocidas como las ecuaciones de TOV, a partir de dichas EdE.
En particular, se estudia la posibilidad de una transición de fases hadrón-quark en el interior de una EH a $\textrm{T}=0$ y se discuten sus consecuencias observacionales.

\section{Modelos}
\label{Modelos}
Para modelar las estrellas de neutrones con núcleos superconductores de color, describiremos la fase hadrónica con EdE modernas que cumplen con las restricciones astrofísicas impuestas por los púlsares de 2~$\textrm{M}_{\odot}$ y el evento GW170817. Estas serán obtenidas a partir del modelo \textit{Walecka} no-lineal con constantes de acoplamiento dependientes de la densidad \citep{2019PhRvC.100a5803M}.

Para la fase de quarks utilizaremos el \textit{Field Correlator Method} (FCM), incluyendo interacciones vectoriales \citep{Klahn:2015mfa} y pares de quarks formando una fase superconductora de color conocida como \textit{Color Flavor Locked} (CFL) \citep{2002PhRvD..66g4017L}.
El FCM es un modelo efectivo de la QCD para el tratamiento de la materia de quarks. La EdE del FCM está parametrizada por dos cantidades: el condensado de gluones, $\textrm{G}_{2}$, y el potencial estático quark-antiquark para largas distancias, $\textrm{V}_{1}$  \citep{2017A&A...601A..21M}. Dado que a bajas temperaturas y altas densidades $\textrm{G}_{2}$ y $\textrm{V}_{1}$ no están bien determinados, en nuestro estudio los consideramos parámetros libres, teniendo en cuenta trabajos anteriores que acotan a $\textrm{V}_{1}$ entre 10~MeV y  100~MeV \citep{2014JPhCS.527a2021L} y  $\textrm{G}_{2}$~=~0.012~$\textrm{GeV}^{4}$ con un 50~\% de incerteza \citep{Burgio:2016cmr}.
Derivamos las cantidades termodinámicas para la materia de quarks a partir del gran potencial $\Omega_{q}$:

\begin{eqnarray}
  \Omega_{q} & = & \frac{3}{\pi^{2}} \left[ \int^{P_{fc}}_{0}p^{2} (p- \tilde{\mu}_{u})  {\,\textrm{d}p} +  \int^{P_{fc}}_{0}p^{2} (p-  \tilde{\mu}_{d}) {\,\textrm{d}p} \right. \nonumber\\
  && +  \left. \int^{P_{fc}}_{0}p^{2}(\sqrt{p^2 + {m_s^2}}- \tilde{{\mu_{s}}}) {\,\textrm{d}p} \right] - \frac{3}{\pi^{2}} \Delta^2 {\overline{\mu}_{*}}^2 \nonumber\\
  &&  -  \frac{\textrm{G}_{\textrm{v}}}{2} ({\textrm{w}_{u}}^2+ {\textrm{w}_{d}}^2+{\textrm{w}_{s}}^2) + \frac{9}{64} {\textrm{G}_2}, 
\label{eq:granpot} 
\end{eqnarray}

\begin{equation}
  \overline{\mu}_{*} = \frac{1}{3} \sum_{i}  \tilde{\mu}_{i},
\end{equation}
  
\begin{equation}
  \tilde{\mu_{i}} = {\mu_{i}} - \frac{1}{2} V_1 - G_{\textrm{v}} \textrm{w}_{i} ,\\ \textrm{con} \ i= u,d,s
\end{equation}
donde $\textrm{w}_{u}$, $\textrm{w}_{d}$ y $\textrm{w}_{s}$ son los campos vectoriales y $\textrm{G}_{\textrm{v}}$, la constante de acoplamiento vectorial, determina la intensidad de la interacción vectorial repulsiva entre los quarks, la cual produce un endurecimiento de la EdE tornando menos compresible la materia de quarks. El parámetro $\Delta$, conocido como gap de la fase superconductora, da cuenta de la intensidad de la fuerza de apareamiento entre los quarks de diferentes sabores y colores que forman pares en la fase CFL.

Los pares de quarks de la fase CFL poseen el mismo momento común de Fermi, $P_{fc}$, que se determina minimizando \ref{eq:granpot}. Esto reduce la energía total del sistema y hace que una fase superconductora de color a altas densidades,  sea más favorable desde el punto de vista energético que un plasma de gluones y quarks libres. Por lo tanto, es posible que la materia de quarks se encuentre en un estado superconductor de color en el núcleo de EHs, {sometida} a densidades varias veces mayores a la densidad de saturación nuclear, $\rho_{0}= 2.7$~x~$10^{14}~\textrm{gr/cm}^{3}$.

\section{Resultados}
\label{Resultados}

\begin{figure}[!t]
  \centering
  \includegraphics[width=0.44\textwidth]{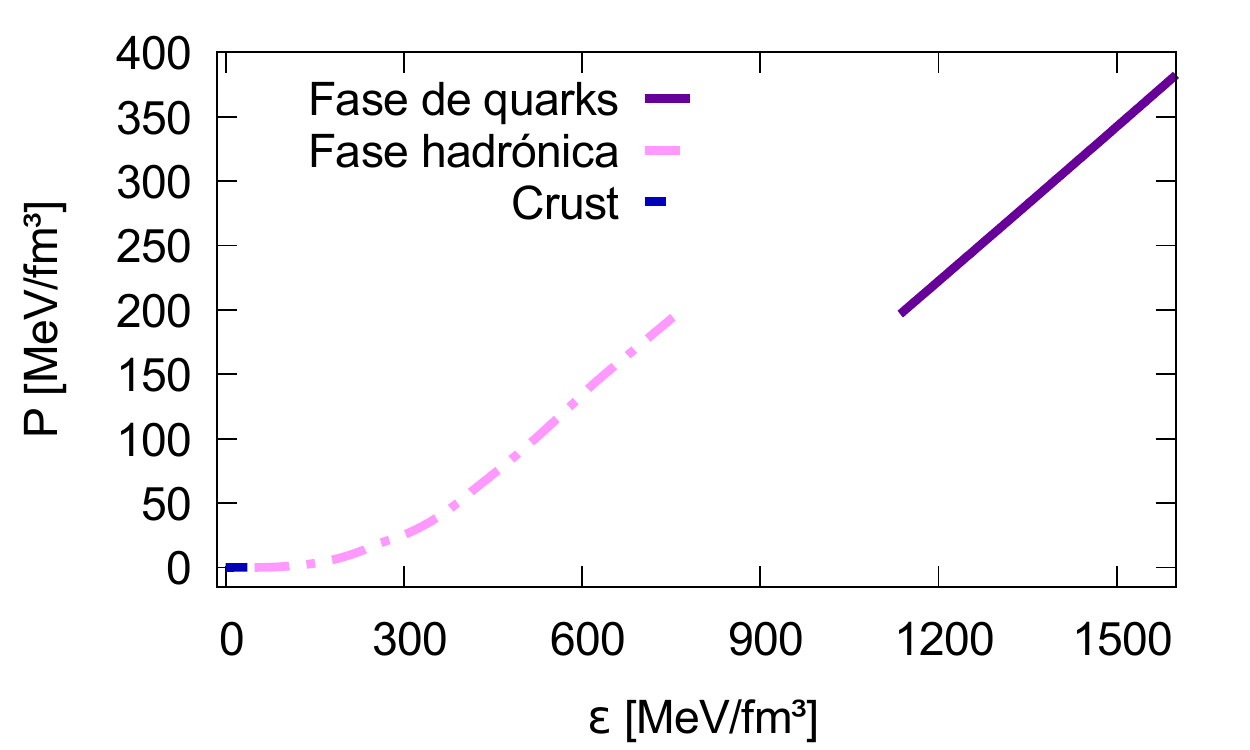}
  \caption{Ecuación de estado híbrida construida a partir de los modelos descriptos y usando el formalismo de Maxwell.}
  \label{fig: EoS}
\end{figure}

Para los parámetros del modelo {(Sec. \ref{Modelos})} elegimos $\textrm{G}_{2}$~$=$~0.001~$\textrm{GeV}^{4}$ y $\textrm{V}_{1}~$=~50~MeV, con lo cual {se obtiene} la EdE que se muestra en la (Fig: \ref{fig: EoS}), donde el gap $\Delta$~=~100~MeV y la constante de acoplamiento vectorial de los quarks $\textrm{G}_{\textrm{v}}$~=~0.007~$\textrm{MeV}^{-2}$. El salto en la densidad de energía $\varepsilon$ corresponde a la transición de fase hadron-quark, con el formalismo de Maxwell. La misma ocurre cuando las presiones en función del potencial químico para la fase hadrónica y la de quarks son iguales.

\begin{figure}[!t]
  \centering
   \includegraphics[width=0.44\textwidth]{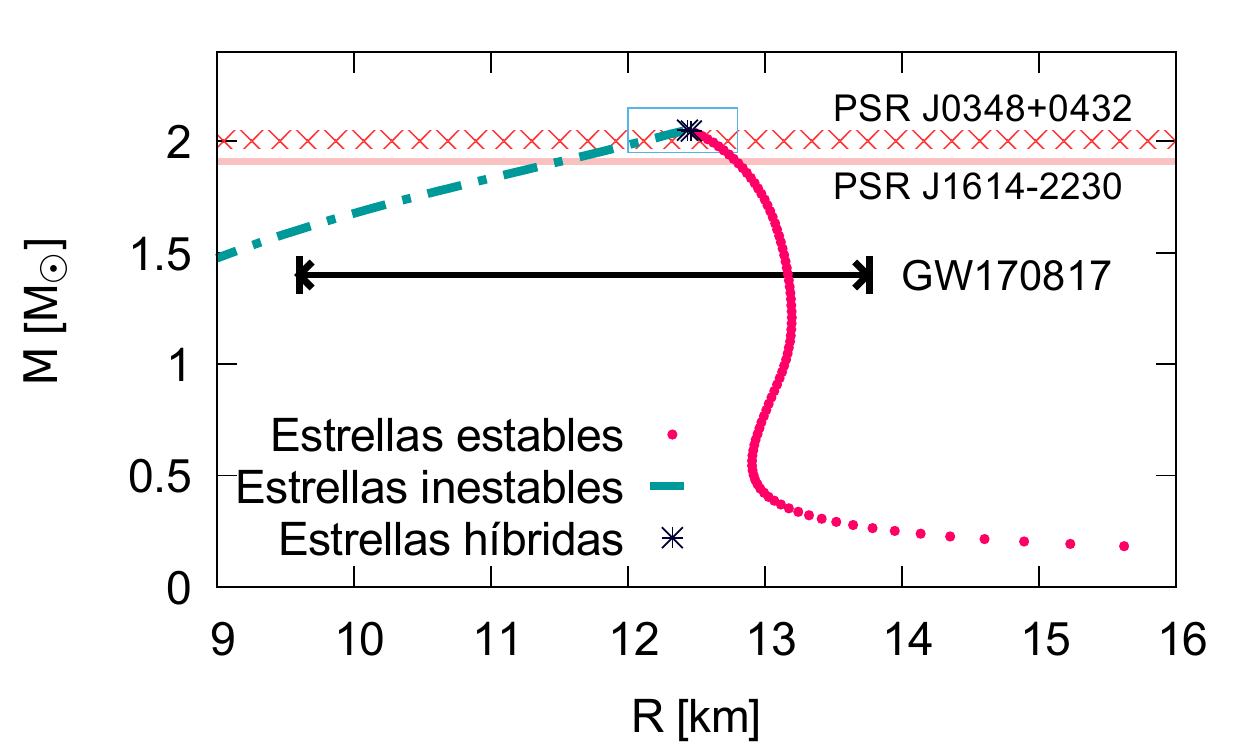}
    \includegraphics[width=0.44\textwidth]{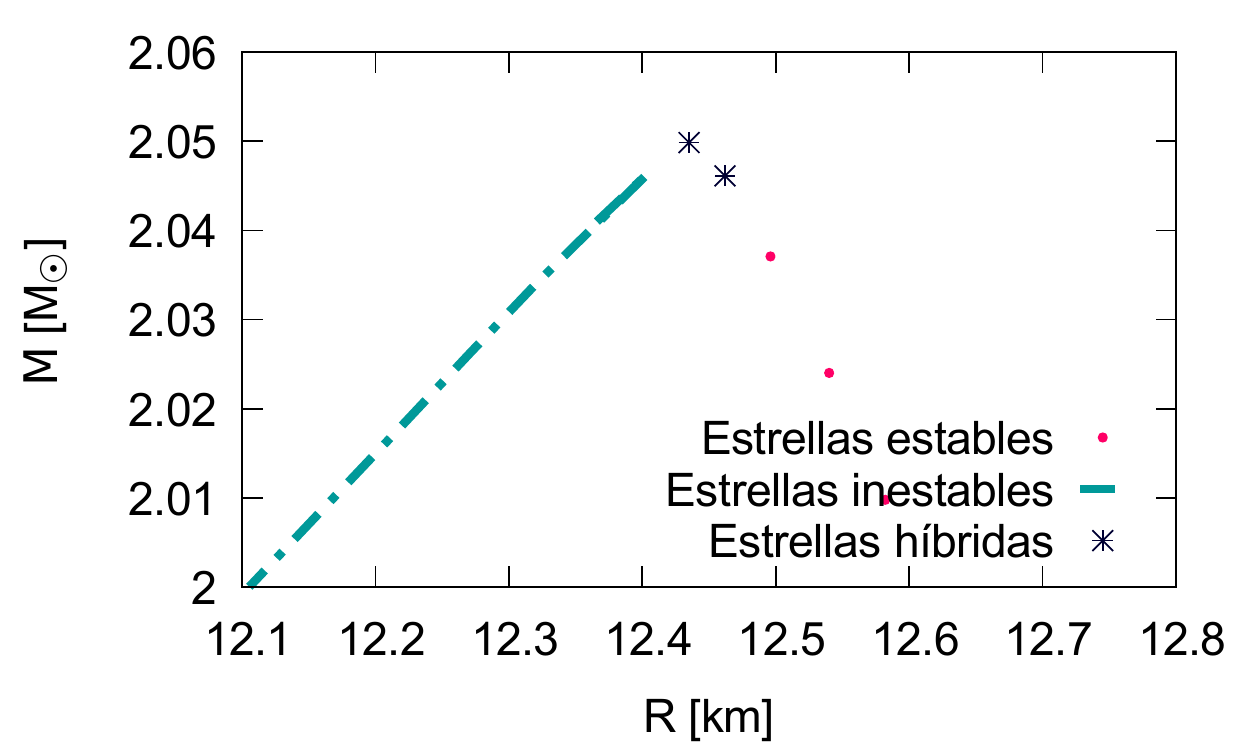}
\caption{{\em Panel superior:} Relación masa-radio para la familia de estrellas construidas, estables e inestables. El asterisco, indica las últimas estrellas de la rama de estabilidad, siendo además las únicas híbridas con núcleo superconductor de color. La línea negra sólida muestra la restricción de radios a las EN que se desprende del evento GW170817. {{\em Panel inferior:} detalle del recuadro del panel superior}.}
\label{fig: m-radio}
\end{figure}

\begin{figure}[!t]  
  \centering
    \includegraphics[width=0.44\textwidth]{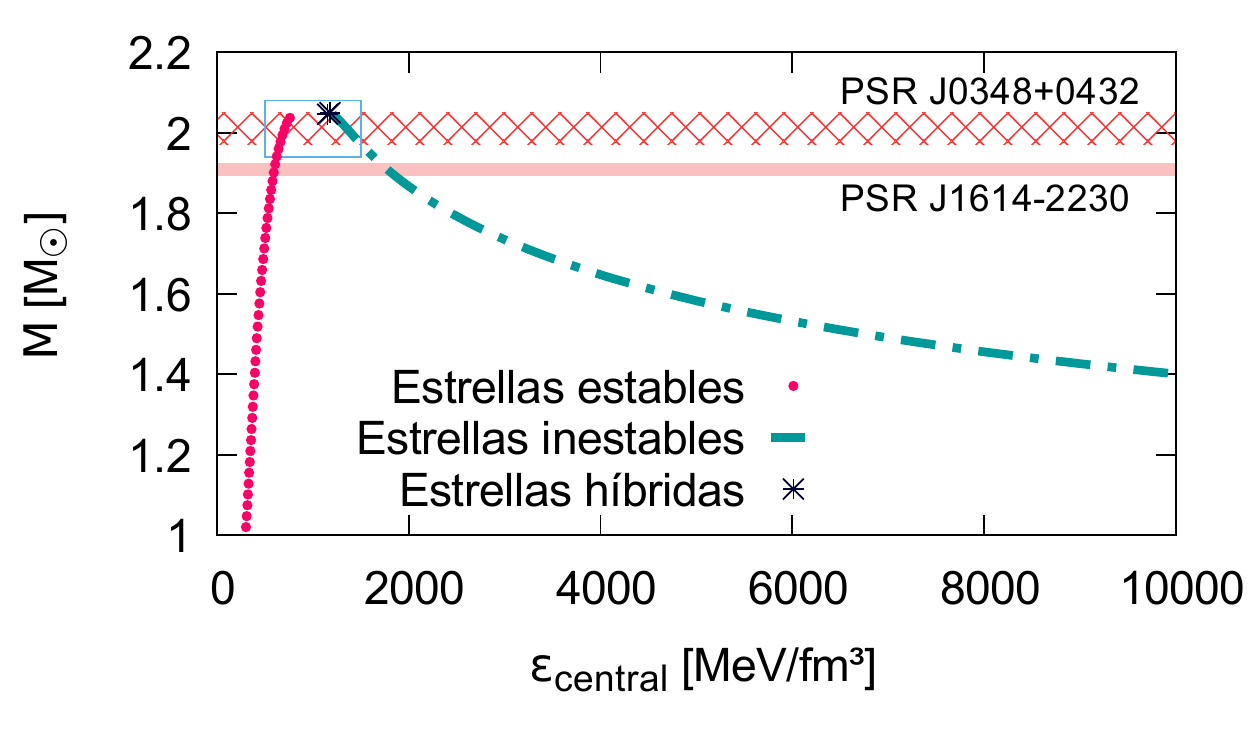}
    \includegraphics[width=0.44\textwidth]{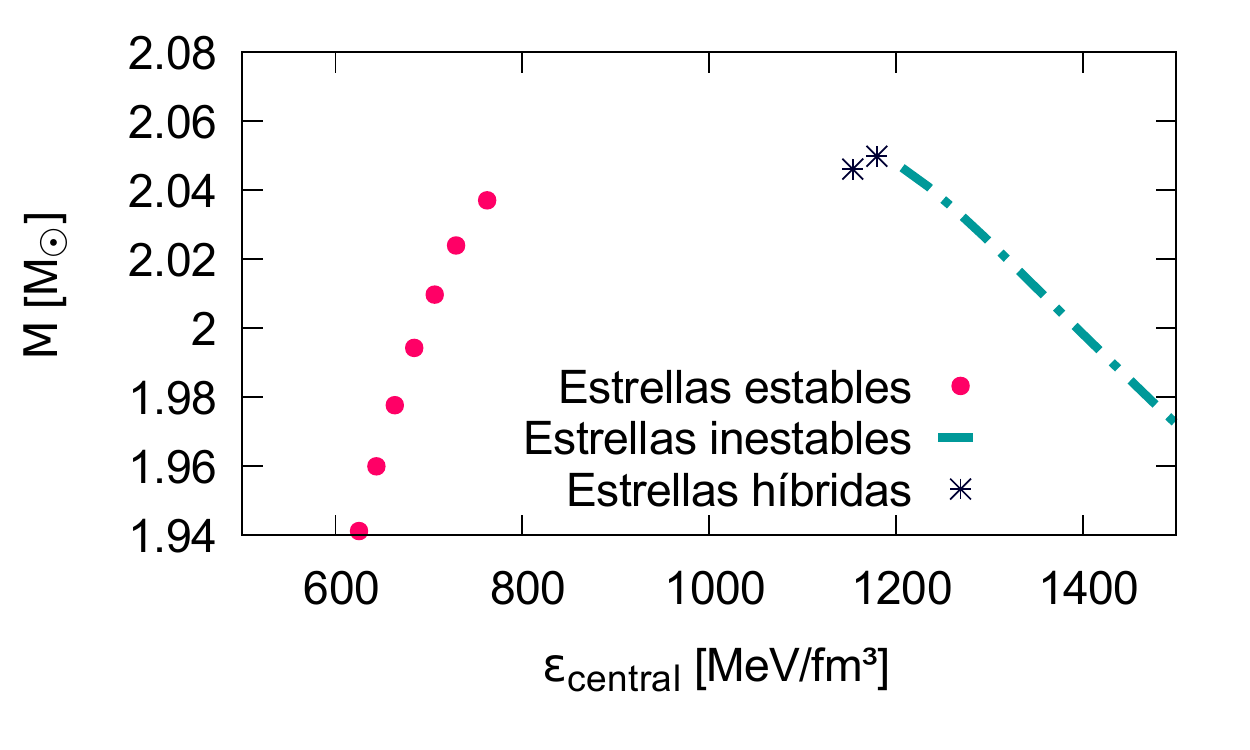} 
\caption{{\em Panel superior:} Relación masa-densidad de energía central para la familia de estrellas. El salto en la densidad es un resultado de la transición de fase abrupta entre la materia hadrónica y la materia de quarks superconductora de color. {{\em Panel inferior:} detalle del recuadro del panel superior}.}
\label{fig: mrho}
\end{figure}

A partir de las EdE híbridas obtenidas resolvimos las ecuaciones relativistas de equilibrio hidrostático derivadas por Tolman, Oppenheimer y Volkoff (TOV) en 1939, que nos permitieron construir familias de estrellas estables y encontrar relaciones tales como masa-radio (Fig: \ref{fig: m-radio}) y densidad de energía central-radio (Fig: \ref{fig: mrho}). 

De la relación masa-radio, se puede constatar que la rama de estrellas estables cumple la condición de una masa máxima mínima de 2~$\textrm{M}_{\odot}$. Además, para las estrellas de 1.4~$\textrm{M}_{\odot}$ se satisface la restricción para los radios impuesta por el evento GW170817 (línea sólida color negra). Las últimas estrellas de la rama estable, entre las dos indicadas con asteriscos azules en las (Fig: \ref{fig: m-radio}) y (Fig: \ref{fig: mrho}) son las únicas estrellas híbridas estables con núcleos superconductores de color, mientras que todas las demás son puramente hadrónicas (sin núcleo de quarks en fase CFL). Asimismo, en la (Fig: \ref{fig: mrho}) se puede ver que efectivamente, luego de la transición de fase, existe una rama estable corta de estrellas híbridas. Además se puede observar un salto en la densidad, correspondiente a la transición de fase aprupta de la EdE.

\begin{figure}[!t]  
  \centering
    \includegraphics[width=0.44\textwidth]{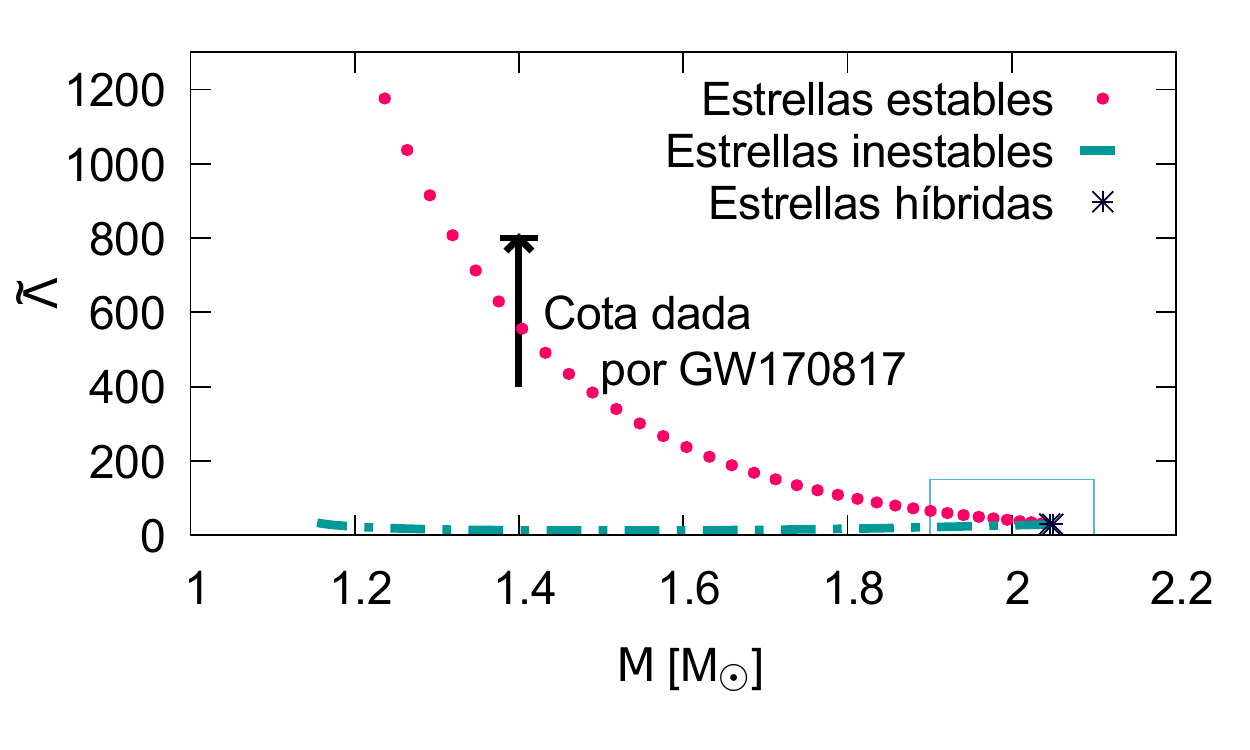}
    \includegraphics[width=0.44\textwidth]{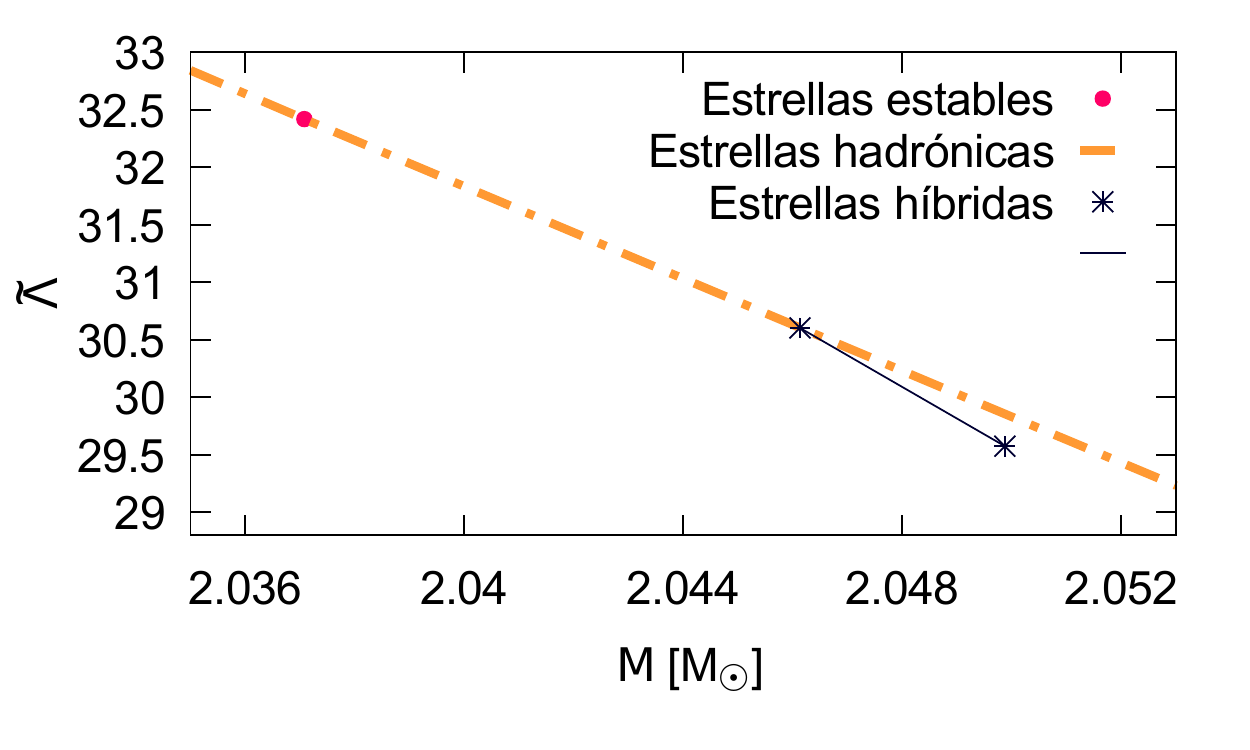} 
\caption{{\em Panel superior:} Deformación {por mareas} adimensional vs. masa gravitacional. {{\em Panel inferior:} detalle del recuadro del panel superior}, que muestra la comparación de las EHs con las puramente hadrónicas.}
\label{fig: tidal}
\end{figure}

Finalmente, es posible calcular la deformabilidad por mareas (\textit{tidal deformability}), $\Lambda$, como función de la masa gravitacional de las ENs, como se muestra en la (Fig: \ref{fig: tidal}).  En la zona ampliada del recuadro se muestran la curva puramente hadrónica y las estrellas estables con núcleos superconductores de color. La presencia de la fase CFL, {disminuye $\Lambda$} si se comparan las estrellas híbridas con las puramente hadrónicas de igual masa gravitacional.

\section{Resumen y conclusiones}
\label{Conclusiones}

En este trabajo fijamos parámetros para el modelado en la fase de quarks (FCM, con interacciones vectoriales y superconductividad de color) y obtuvimos una EdE híbrida. {De} las ecuaciones de TOV encontramos la rama de estrellas estables y las relaciones que las describen. 

La masa {gravitacional} máxima de una EN es {la que} puede tener una estrella antes de volverse inestable y colapsar a un agujero negro \citep{1996A&A...305..871B}, donde la condición de estabilidad viene dada por dM/$\textrm{d}\varepsilon_{c}$~$>$~0. 

{Nuestros resultados indican} que las {ENs estables} obtenidas verifican las cotas observacionales de masa y radios impuestas por los púlsares {de $\sim 2\textrm{M}_{\odot}$} y por los cálculos de deformabilidad de marea obtenidos a partir del evento GW170817. {Hallamos que la rama de EHs es muy corta debido} a que la materia de quarks tiende a desestabilizar a la EN \citep{Mariani:2017syz}.

{En trabajos futuros modificaremos} los valores de los cuatro parámetros {libres} $\textrm{G}_{2}$, $\textrm{V}_{1}$, $\textrm{G}_{\textrm{v}}$ y $\Delta$ para la fase de quarks, ampliando la búsqueda de diferentes combinaciones en el espacio de parámetros, estudiar si es posible obtener una rama de EHs estables más extendida.

{Planeamos} incluir en los modelos estudiados el efecto de campos magnéticos intensos, {relevante al estudio de ENs. Mediante la extensión de los modelos utilizados a temperatura finita, esperamos} obtener las EdE para modelar la materia del interior de proto-estrellas híbridas y fusiones (\textit{mergers}) de ENs.

{Analizaremos} el espectro de modos de oscilación no radiales de estos objetos, los cuales generan ondas gravitacionales. {En particular}, el modo de gravedad {$g$,} asociado a alguna discontinuidad en la EdE, podría {aportar} información relevante de la materia que compone las ENs. {Finalmente}, exploraremos la posible existencia de una fase mixta utilizando el formalismo de Gibbs que permitirá comparar ambos formalismos (Maxwell y Gibbs) para el tratamiento de la transición de fase.

\begin{acknowledgement}
{Agradecemos al} CONICET y {la} UNLP por {su} apoyo financiero, en el marco de los proyectos PIP-0714, y G157 y X824, respectivamente. {D.C.} agradece al COL por la ayuda económica para asistir a la reunión. {Agradecemos} a M. Mariani {por sus valiosas sugerencias y comentarios}.

\end{acknowledgement}


\bibliographystyle{baaa}
\small
\bibliography{bibliografia}
 
\end{document}